\pdfoutput=1

\documentclass[aps,superscriptaddress,twocolumn,nofootinbib]{revtex4}
\usepackage{natbib}
\usepackage{dcolumn}
\usepackage{graphics}
\usepackage{subfigure}
\usepackage{amsmath}
\usepackage{amssymb, amsfonts}
\usepackage[official]{eurosym}
\usepackage{color}
\begin{document}
\def\refname{References and Notes}

\makeatother

\title{The non-random walk of stock prices:\\  The long-term correlation between signs and sizes}

\author{Gabriele La Spada}
\email{glaspada@luiss.it}
\affiliation{LUISS Guido Carli University, Viale Pola 12, 00198 Roma, Italy}
\affiliation{Santa Fe Institute, 1399 Hyde Park Road, Santa Fe, NM 87501}

\author{J. Doyne Farmer}
\affiliation{Santa Fe Institute, 1399 Hyde Park Road, Santa Fe, NM 87501}
\affiliation{LUISS Guido Carli University, Viale Pola 12, 00198 Roma, Italy}

\author{Fabrizio Lillo}
\affiliation{Santa Fe Institute, 1399 Hyde Park Road, Santa Fe, NM 87501}
\affiliation{INFM Unit\`a di Palermo and Dipartimento di Fisica e Tecnologie Relative, viale delle Scienze I-90128, Palermo, Italy}

\date{\today}
\begin{abstract}
We investigate the random walk of prices by developing a simple model relating the properties of the signs and absolute values of individual price changes to the diffusion rate (volatility) of prices at longer time scales. We show that this benchmark model is unable to reproduce the diffusion properties of real prices. Specifically, we find that for one hour intervals this model consistently over-predicts the volatility of real price series by about $70\%$, and that this effect becomes stronger as the length of the intervals increases.  By selectively shuffling some components of the data while preserving others we are able to show that this discrepancy is caused by a subtle but long-range non-contemporaneous correlation between the signs and sizes of individual returns. We conjecture that this is related to the long-memory of transaction signs and the need to enforce market efficiency. 
\end{abstract}
\pacs{
        {89.65.Gh} {Economics; econophysics, financial markets, business and management} \and
        {05.40.Jc} {Brownian motion} \and
        {02.50.Ey} {Stochastic processes}
        } 
\maketitle
\section{Introduction}
\label{intro}

The random walk was originally introduced in finance in 1900 \cite{Bachelier64} as an empirical model for prices.  It is still widely used in finance for many practical problems, such as option and interest rate pricing. The conceptual justification for the random walk description of asset price is market efficiency, i.e. that asset price changes should be unpredictable \cite{Cootner64}.  Even if prices do not follow a perfect random walk, for many purposes this is an excellent approximation:  While there may be some structure in the drift term, so that occasionally clever arbitrageurs can predict and exploit small deviations from randomness, basically the direction of price movements is very close to random.

The well-known non-random exception is the diffusion rate of prices, which in finance is usually called the {\it volatility}.  As first carefully documented by Engle \cite{Engle82}, volatility varies in a way that is quite persistent in time.  Its autocorrelation function dies out slowly with an asymptotic power law decay for long times, so that it is a long-memory process \cite{Ding93,Breidt93,Harvey93,Montero05}.  The question of what causes variations in volatility is more complicated, and remains unanswered.  Standard equilibrium theories in economics say that volatility should be caused by new information \cite{Campbell97}, but new information is difficult to measure, and while a few recent studies seem to support this on longer time scales \cite{Engle05,Engle06}, there are several studies on shorter time scales suggesting that the correlation between volatility and news is weak \cite{Roll84b,French86,Cutler89}.  

An alternative approach is to look for immediate causes of volatility.  For example, Clark suggested modeling volatility as a subordinated stochastic process, in which the transaction rate varies\footnote{Variations in the transaction rate may or may not depend on information, thus this idea neither supports nor contradicts the standard theory.} and consequently the diffusion rate also varies \cite{Clark73}.  Variations in the transaction rate correlate with volatility, so this theory is at least partially correct \cite{Plerou00,Ane00,Anderson03}.  However, a more recent study shows that, at least on a time scale of fifteen minutes, this is not the dominant correlate of volatility.  Instead, there is a much larger effect due to the size of individual price changes, which is only weakly correlated with the size of transactions and with the transaction frequency \cite{Gillemot05}.   The long-memory properties of individual price changes also match those of volatility much better than those of volume or transaction frequency.  

This story is further complicated by the fact that transaction signs have long-memory \cite{Bouchaud04,Lillo03c,Bouchaud04b,Farmer06}.  Transactions can be labeled as having a positive sign if they are initiated by a buyer, i.e. if the trading order that triggers the transaction is from a buyer, and similarly as having a negative sign if they are initiated by a seller.  Sequences of transaction signs have long-memory, i.e. their autocorrelation functions $C(\tau)$ decay as a power law $C(\tau) \sim \tau^{-\gamma}$ with $\gamma < 1$, typically with $\gamma \approx 0.5$.  This strong autocorrelation structure implies that the signs of transactions are quite predictable using a trivial algorithm.  Since buyer initiated transactions tend to push the price up and seller initiated transaction tend to push it down, this suggests that prices should also be predictable, which would contradict market efficiency.  To prevent this from happening there must be a non-trivial relationship between transactions and price responses.

We add to this story by studying a simple model for the aggregation properties of non-zero price returns at the level of individual transactions.  In particular, we view price changes as the steps in a generalized random walk.  The term {\it generalized random walk} refers to the possibility that there are correlations in the signs of the steps and their sizes.  Under the assumption that price changes are permanent, we develop a model predicting the expected volatility in terms of properties of the generalized random walk, such as the number of steps, the average step size, the variance of the step sizes, the imbalances between positive and negative steps, and sums of the autocorrelation functions for step signs and sizes.  Restated in terms of prices, the model is based on the number of non-zero price changes, the average size of price changes, the variance of the size of price changes, the imbalance between up and down price movements, and the sum of the autocorrelation functions of price change sign and size.  We show that this model performs poorly in describing the volatility of real data.
We show that the predictions of this model for volatility are much too large, by the order of $70\%$ of the empirical volatility, and that the cause of this over-prediction is the relationship between lagged signs and sizes of the price changes.


The paper is organized as follows. In Section~\ref{GRWModel} we develop a model for the random walk of prices, and in Section~\ref{DataSet} we describe the data.   In Section~\ref{TestingTheHypotheses} we test the hypotheses of the model, and in Section~\ref{VolatilityEstimation} we present our main empirical results. Finally, Section~\ref{Conclusions} we summarize the paper, discuss what the results mean, and outline future work.

\section{The generalized random walk}\label{GRWModel}

Price returns are defined in logarithmic terms as $r_t = \log p_t - \log p_{t-1}$, where $p_t$ is the price at time $t$.  In this paper we will take $p_t$ to be the {\it midprice}, i.e. the average of the best quoted buying and selling prices.   We consider only non-zero returns $r_t \neq 0$, and define the time variable $t$ under the transformation $t \to t+1$, which occurs whenever the midprice changes.  That is, except where otherwise stated, throughout this paper time is just a counter labeling the number of non-zero steps for the random walk of price changes.

An additive stochastic process 
\begin{equation}
R_n \equiv \sum_{t=1}^n r_t,
\end{equation}
where the increments $r_t$ are stationary, defines what we will call a {\it generalized random walk}.  We use this term to distinguish it from a ``pure" random walk, in which the increments $r_t$ are independent and identically distributed (IID).  Our purpose here is to make a model for the squared volatility, which we will measure as the variance $Var ( R_n )$, in terms of the underlying properties of a generalized random walk with increments $r_t$.   For this purpose it is useful to decompose the individual returns as $r_t = s_t w_t$, where $s_t$ is the sign of the return at time $t$, and $w_t$ is its magnitude.
 
 \subsection{Assumptions \label{assumptions}}
 
To make our analysis tractable we make the following assumptions:
\begin{itemize}
\item[$(i)$] $s_t$ is a Bernoulli variable $\forall \,t$.
\item[$(ii)$] $w_t$ is a strictly positive random variable $\forall \, t$.
\item[$(iii)$] Both $\{s_t\}$ and $\{w_t\}$ are wide sense stationary processes\footnote{A stochastic process $\{X_t\}$, where $t = $ represents the integers, is  wide sense stationary (WSS)  if ($i$) $\mathbb{E}\left[X_t^2\right] < \infty \quad \forall \, t$, ($ii$) $\mathbb{E}\left[X_t\right] = \mu_X \quad \forall \, t$, ($iii$) $\mathbb{E}\left[X_tX_u\right] = \mathbb{E}\left[X_{t+h}X_{u+h}\right] \quad \forall \, t, u, h$.}.
\item[$(iv)$]  $\{s_t\}$ and $\{w_t\}$ are independent stochastic processes.
\end{itemize}
The first two assumptions are simply the decomposition of any single step into its sign and its magnitude.  The third assumption of stationarity is important because it implies that autocorrelation function $c_X(t,u)$ of the process $X$ between times $t$ and $u$ only depends only on the lag, i.e. $c_X(t,u) = c_X(|t-u|)$. Note that in making this assumption we are also assuming that the first two central moments of the distribution of $w_t$ are finite (this is automatic for $s_t$). The fourth hypothesis greatly simplifies calculations, since under the independence hypothesis the joint probability density function of any given subset of these variables can be factorized.  We will see that the first three assumptions are fine, but the fourth assumption is not well-satisfied for the real data.

\subsection{Derivation of formula for volatility}

The squared volatility is the variance of $R_n$ and can be computed as %
\begin{eqnarray}
Var(R_n) &\equiv& \mathbb{E}[R_n^2]-\mathbb{E}[R_n]^2\nonumber\\
&=& \mathbb{E}\left[\sum_{i=1}^n s_iw_i \sum_{j=1}^n s_jw_j \right] - \mathbb{E}\left[\sum_{i=1}^n s_iw_i\right]^2\nonumber\\
&=& \sum_{i}^n\mathbb{E}\left[w_i^2\right]+2\sum_{i<j}^n\mathbb{E}\left[s_is_jw_iw_j\right] - \left(\sum_{i}^n\mathbb{E}[s_iw_i]\right)^2,\nonumber
\end{eqnarray}
where  $\mathbb{E}[.]$ represents the expected value.


Let the the means of $s$ and $w$ be $\mu_s$ and $\mu_w$ and the variances be $\sigma_s^2$ and $\sigma_w^2$.   Since both the $s$-process and the $w$-process are stationary we can write $\mathbb{E}[s_i]=\mu_s$, $\mathbb{E}[w_i]=\mu_w$, and $\mathbb{E}[w_i^2]=\sigma_w^2+\mu_w^2$ for all $i$.   Moreover $w_t$ and $s_t$ are independent processes and we can factorize $\mathbb{E}[s_iw_i]=\mathbb{E}[s_i]\mathbb{E}[w_i]$ and $\mathbb{E}\left[s_is_jw_iw_j\right] = \mathbb{E}\left[s_is_j\right]\mathbb{E}\left[w_iw_j\right]$. Then we obtain 

 \begin{eqnarray}
 \nonumber
Var(R_n) = & & ~~~~~~~~~~~~~~~~~~~~~~~~~~~~~~~~~~~~~~~~~~~~~~~~~~~~~~~~~~~ \nonumber\\
\sum_i^n (\sigma_w^2 +\mu_w^2)  &+&  2\sum_{i<j}^n\mathbb{E}\left[s_is_j\right]\mathbb{E}\left[w_iw_j\right] -  \left(\sum_{i}^n\mu_s\mu_w\right)^2\nonumber\\
=  n[\sigma_w^2+\mu_w^2] 
& + & 2 \sum_{i<j}^n \left[c_s(i, j)\sigma_s^2 + \mu_s^2\right]\left[c_w(i, j)\sigma_w^2 + \mu_w^2\right]\nonumber\\
 - n^2\mu_s^2\mu_w^2,\nonumber
 \label{GRWVar_0}
 \end{eqnarray}
where $c_s(i, j)$ and $c_w(i, j)$ are the autocorrelation functions of the sign process and the size process.
The sum in the second term can be written explicitly as
\begin{eqnarray}
\sum_{i<j}^n && \left[c_s(i, j)\sigma_s^2 + \mu_s^2\right]\left[c_w(i, j)\sigma_w^2 + \mu_w^2\right] =\nonumber\\
&=&\left(\sigma_s^2\sum_{i<j}^n c_s(i, j)c_w(i, j) + \mu_s^2 \sum_{i<j}^n c_w(i, j)\right) \sigma_w^2 +\nonumber\\
&&+\left(\sigma_s^2 \sum_{i<j}^n c_s(i, j) + \sum_{i<j}^n\mu_s^2\right) \mu_w^2\,.
\label{GRWVar_1}
\end{eqnarray}
Let $f: \mathbb{N}\times\mathbb{N} \rightarrow \mathbb{R}$ be a generic function of two integer variables. If $f(i, j) = f(|i-j|)$ for all $i$ and $j$, then
\begin{equation}
\sum_{i<j}^n f(i, j) = n \sum_{l=1}^n \left(1 - \frac{l}{n}\right)f(l)\,.
\label{SumAcfStationary}
\end{equation}
Since both $s_t$ and $w_t$ are stationary, Eq.~\ref{SumAcfStationary} holds both for $c_s$ and $c_w$. Then, we can use (\ref{SumAcfStationary}) in (\ref{GRWVar_1}) and Eq.~(\ref{GRWVar_0}) becomes
\begin{eqnarray}\label{GRWVarFinal}
\hat{V} & \equiv & Var(R_n) = n\Big\{\left[1 + 2 \sigma_s^2 K_{s,w}(n) + 2 \mu_s^2 K_{w}(n)\right] \sigma_w^2 + \nonumber\\
&& + \left[1+ 2 \sigma_s^2 K_s(n) - \mu_s^2\right]\mu_w^2\Big\}\,,
\end{eqnarray}
where $K_{s,w}(n) = \sum_{l=1}^{n}\left(1-\frac{l}{n}\right) c_s(l)c_w(l)$, $K_{w}(n) = \\\sum_{l=1}^{n}\left(1-\frac{l}{n}\right) c_s(l)$ and $K_{s}(n) = \sum_{l=1}^{n}\left(1-\frac{l}{n}\right) c_s(l)$ are functions of the number of steps $n$ and depend on the autocorrelation structures of both signs and sizes.  We have introduced the notation $\hat{V}$ to emphasize that we will be using this as a prediction for squared volatility.



\section{Data}\label{DataSet}
To test the validity of Eq.~(\ref{GRWVarFinal}) we used data for four highly capitalized stocks traded in the London Stock Exchange (LSE).  The stocks are Astrazeneca (AZN),  LLoyds Tsb Group (LLOY), Shell Transport \& Trading Co. (SHEL), and Vodafone Group (VOD).
The investigated period spans more than two years ranging from May 2, 2000  to December 31, 2002, for a total of $675$ trading days.  Summary statistics are given in Table~\ref{Summary&TailIndex&HurstExpW_4Stocks}.  The total number of non-zero returns in the sample is roughly $300,000$ for each of these stocks.  There is on average about one price change per minute, but the trading activity fluctuates considerably and in some periods there can be more than a price change per second.

We have left out the first and last fifteen minutes of each trading day. This choice avoids biases due to the extremely high activity at market opening and closing. If we include data for the full day we get practically the same results. Overnight price changes are omitted.  We have also performed tests removing outliers
and found that this makes no difference in our results.

%
%
\begin{table*}[htdp]
\caption{Summary statistics of the stocks in our sample. Sample size, tail index $\alpha$ (Hill estimator) and Hurst exponent H of the absolute returns.  Note that the quoted significance intervals are standard errors and are much too small.}
\begin{center}
\begin{tabular}{|l|c|c|c|c|}
\cline{2-5}  
\multicolumn{1} {l}{} & \multicolumn{1}{|c|}{AZN} & \multicolumn{1}{c|}{LLOY} & \multicolumn{1}{c|} {SHEL} & \multicolumn{1}{c|} {VOD} \\
\cline{2-5}\noalign{\smallskip}
\hline
Number of trades $\times 10^5$ &  5.5  & 5.7 & 5.9 & 9.4\\
\hline
Number of non-zero returns $\times 10^5$ & 3.2  & 2.7 & 2.7 & 3.4\\
\hline
Trades per 15 min. &  23.7  &  24.9 & 25.4 & 43.5\\
\hline
Non-zero returns per 15 min. &  14.8  & 12.4 & 12.6  & 15.6\\
\hline\noalign{\smallskip}
\hline
Tail index of $w_t$ ($\alpha$) &  $3.0 \pm0.04$  & $ 3.2 \pm0.05$ &  $3.7\pm0.05$   & $8.6\pm0.11$\\
\hline
Hurst exponent of $w_t$ ($H$) &  $0.80 \pm0.006$  &  $0.80\pm0.005$ & $0.85\pm0.011$  & $0.86\pm0.014$\\
\hline
\end{tabular}
\end{center}
\label{Summary&TailIndex&HurstExpW_4Stocks}
\end{table*}

\section{Testing the hypotheses of the model}\label{TestingTheHypotheses}

Before testing the model we first test the hypotheses of the model given in Section~\ref{assumptions}.  
In particular we check the stationarity of both $s_t$ and $w_t$, compute their autocorrelation functions, and test their independence.

To test for stationarity we used two standard tests that are widely employed in time series analysis, the augmented Dickey-Fuller test and the Phillips-Perron test \cite{Hamilton94}.  Both are tests for the null hypothesis that a time series $x_t$ has a unit root $a$, i.e. that under the model $x_t = a x_{t-1} + \eta_t$, where $\eta_t$ is IID noise, $a = 1$.   We applied these tests to the entire time series of signs and sizes.  For each stock we found that for both $s_t$ and $w_t$  the null hypothesis of non-stationarity (unit root) can be rejected with a p-value smaller\footnote{Such a strong result is partially  due to the high number of data points in each sample, but we also due to the fact that the computed root is always much smaller than one (about $10^{-2}$) in both cases.} than $2 \times10^{-16}$.  We also applied these tests to individual days of data and found that the null hypothesis of non-stationarity can be rejected with a p-value smaller than $0.05$ in more than the $95\%$ of the days, with essentially the same results for all stocks. We therefore conclude that both $s_t$ and $w_t$ can be considered stationary processes. 

The wide-sense stationarity hypothesis assumes that the first two central moments $\mu_w$ and $\sigma_w^2$ of $w_t$ are finite.  Since these also appear in equation (\ref{GRWVarFinal}) it is particularly important to test that this is true.  Many studies have shown that the probability distribution of price returns have power law tails, i.e. that $\mathbb{P}(|r_t|>x)\sim x^{-\alpha}$ as $x \rightarrow \infty$, with $0 < \alpha < \infty$.  This implies that moments less than $\alpha$ exist, but moments greater than $\alpha$ are infinite.  In particular, $\alpha > 2$ is sufficient to guarantee that $\mu_w$ and $\sigma_w^2$ are both well-defined.  Early studies suggested that price returns are described by L\`evy distributions, which have $\alpha < 2$  \cite{Fama65,Mandelbrot63}, but most later studies have measured $\alpha > 2$ \cite{Longin96,Lux96,Gopikrishnan98}.  Just to make sure, we estimated the tail index using a Hill estimator \cite{Hill75} as presented in Table~\ref{Summary&TailIndex&HurstExpW_4Stocks}.  In every case we find\footnote{For Vodafone we find $\alpha \approx 8.5$, which calls into question whether the returns really obey a power law at all.  In any case this does not matter for our results here.} that $\alpha > 2$.
We conclude that the first and the second moment of the absolute return distribution exist.

We also studied the autocorrelation structure of $s_t$ and $w_t$.  For each stock we estimated autocorrelation functions from the entire sample.  We find that the autocorrelation of the absolute returns  $w_t$ is a long-memory process, i.e., the autocorrelation function is asymptotically a power-law $c_{w}(t)\simeq c_w(0)t^{-\gamma}$ with $\gamma < 1$.  In contrast, the sign process has some non-zero structure in its autocorrelation function, but is not long-memory.

\begin{figure}[htbp!]
\centering
	\resizebox{1.0\columnwidth}{!}{
		\subfigure[Absolute retruns]%
		{\includegraphics{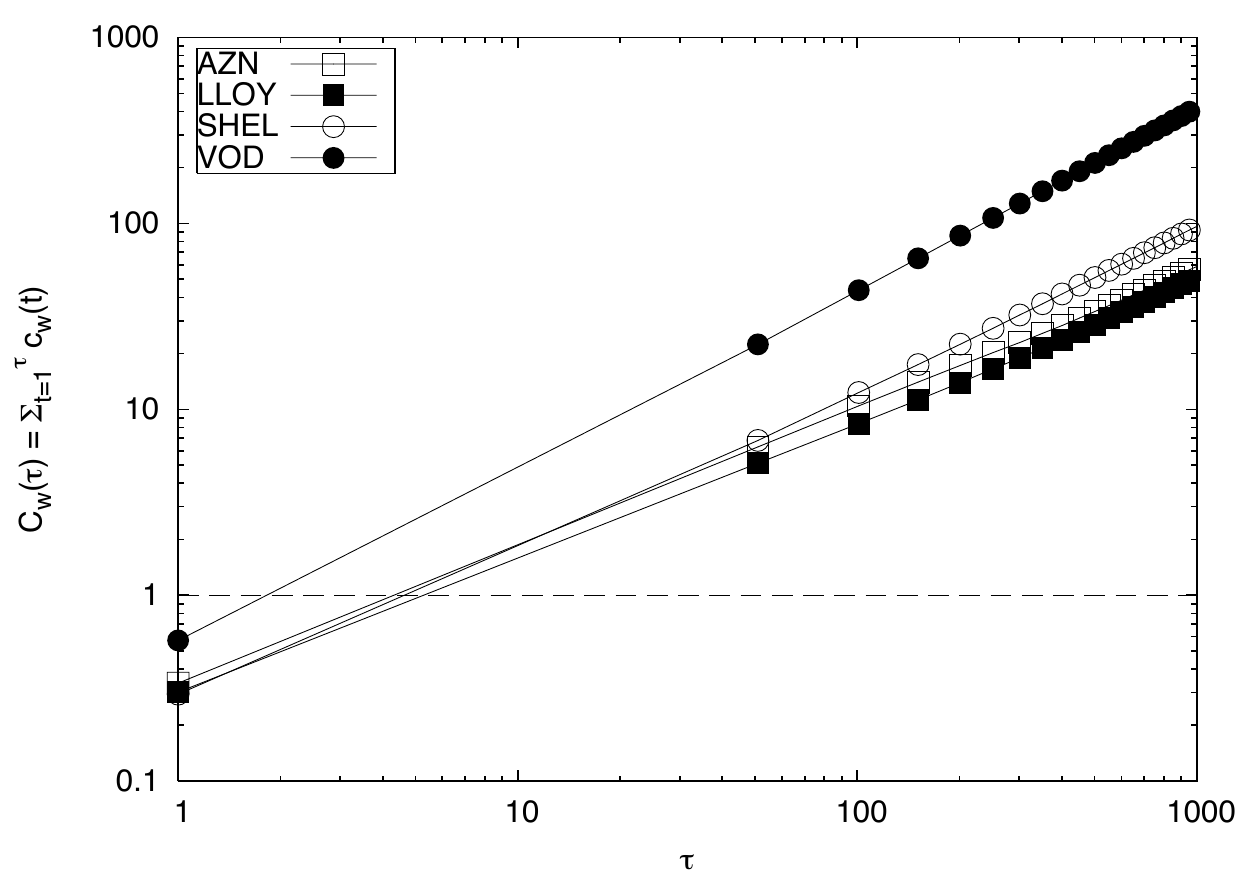}}
	}\\
	\resizebox{1.0\columnwidth}{!}{
		\subfigure[Signs\label{SumAcfS_4Stocks}]%
		{\includegraphics{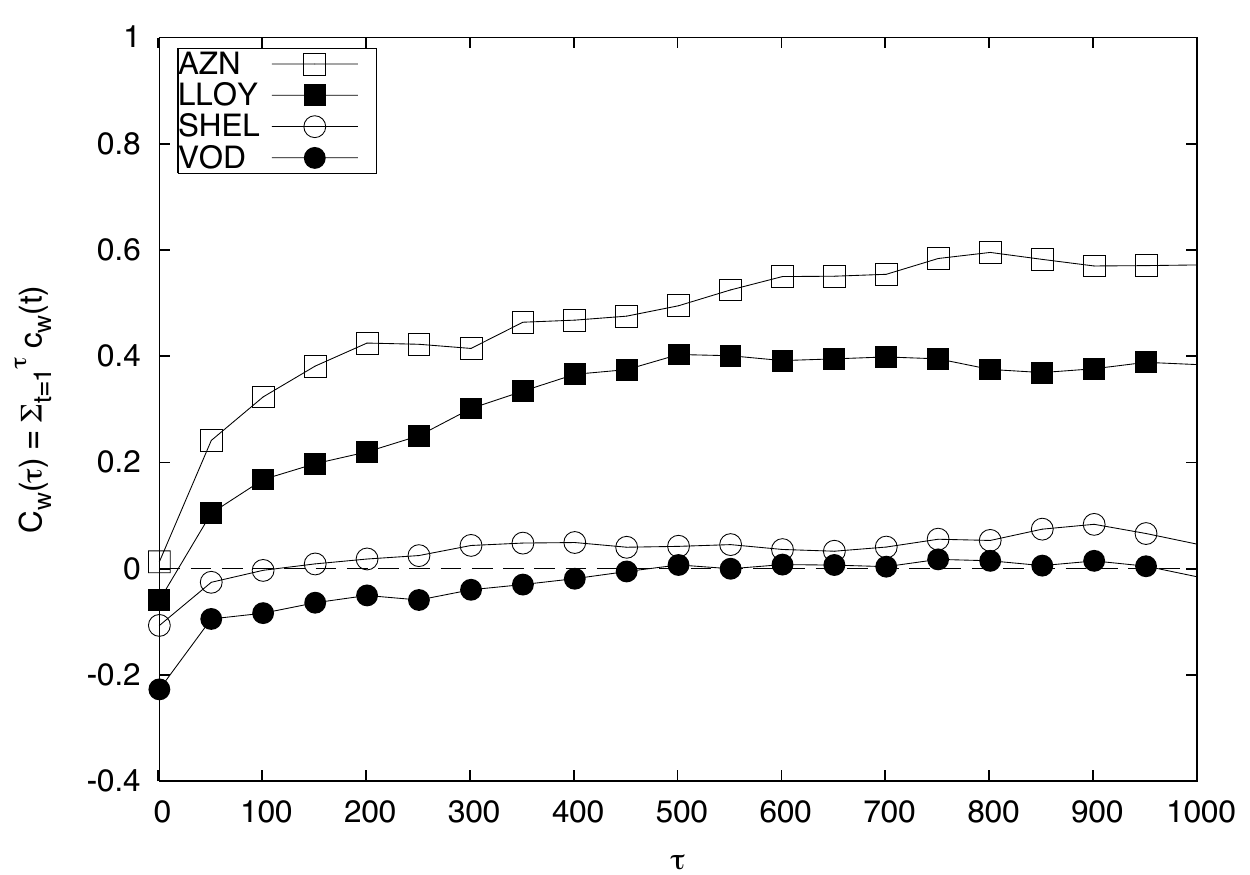}}
	}
\caption{Cumulative autocorrelation functions for the four investigated stocks. Top: Absolute returns in double logarithmic scale.  Bottom: Signs in linear scale.}
\label{SumAcf_4Stocks}
\end{figure}

To give a qualitative feeling for the long-memory nature of the size process, in Fig. (\ref{SumAcf_4Stocks}a) we show $C_w(\tau) \equiv \sum_{t=1}^\tau c_w(t)$, the cumulative sum of the autocorrelation function up to time $\tau$, in double logarithmic scale.  This makes it clear that a power law is a reasonable approximation and that the integral is increasing without bound.  The long-memory nature of a stochastic process can also be characterized by the Hurst exponent $H$, which is related to the decay exponent $\gamma$ as $\gamma=2-2H$ \cite{Beran94}.  From a statistical point of view, computing the Hurst exponent is a more reliable indicator of long-memory than working with the autocorrelation function. We estimate $H$ by using the Detrended Fluctuation Analysis (DFA) introduced in \cite{Peng94}.
In Table~\ref{Summary&TailIndex&HurstExpW_4Stocks} we report the value of the Hurst exponents of $w_t$ for different stocks\footnote{Given the long-memory of the absolute returns we expect that the variance of the absolute return on smaller time scales (e.g. 15 minutes or 1 hour) will be slightly smaller than the variance computed on the entire sample and we indeed observe this.}.  We see that $H$ is always in $[0.80, 0.86]$, which implies $\gamma \in [0.28, 0.40]$. 

In contrast the autocorrelation function of the signs $s_t$ looks completely different. In every case, after at most $10$-$15$ trades the size of the autocorrelation function becomes small enough to be within statistical error.  Even though each coefficient is small, there may be trends in which nearby coefficients tend to be positive or negative, so we have once again computed $C_s(\tau)\equiv\sum_{t=1}^\tau c_s(t)$, as shown in Fig.~(\ref{SumAcf_4Stocks}b). We cannot plot this in double logarithmic scale because of negative values, so we show it in linear scale.  We see that there are some persistent effects involving the accumulation of small terms in the autocorrelation function; in some cases $C_s(\tau)$ takes hundreds of transactions to approach its asymptotic value.  Nevertheless, the behavior is dramatically different from the long-memory behavior of $C_w(\tau)$, as evidenced by the fact that for the signs $C_s(1000) < 0.6$ in every case, whereas for the sizes $C_w(1000) > 40$, and in some cases is closer to $400$. 

In conclusion absolute returns are persistent in time, in agreement with other studies that have found that volatility is a long-memory process \cite{Ding93,Breidt93,Harvey93,Montero05}.  On the contrary signs are weakly autocorrelated, which they have to be to be compatible with market efficiency.  The signs of non-zero returns should not be confused with the signs of transactions, which as we have already mentioned form a long memory process \cite{Bouchaud04,Lillo03c}.  We say more about the possible importance of this in the conclusions.

\begin{figure}[htbp!]
\resizebox{1.0\columnwidth}{!}{
	\includegraphics{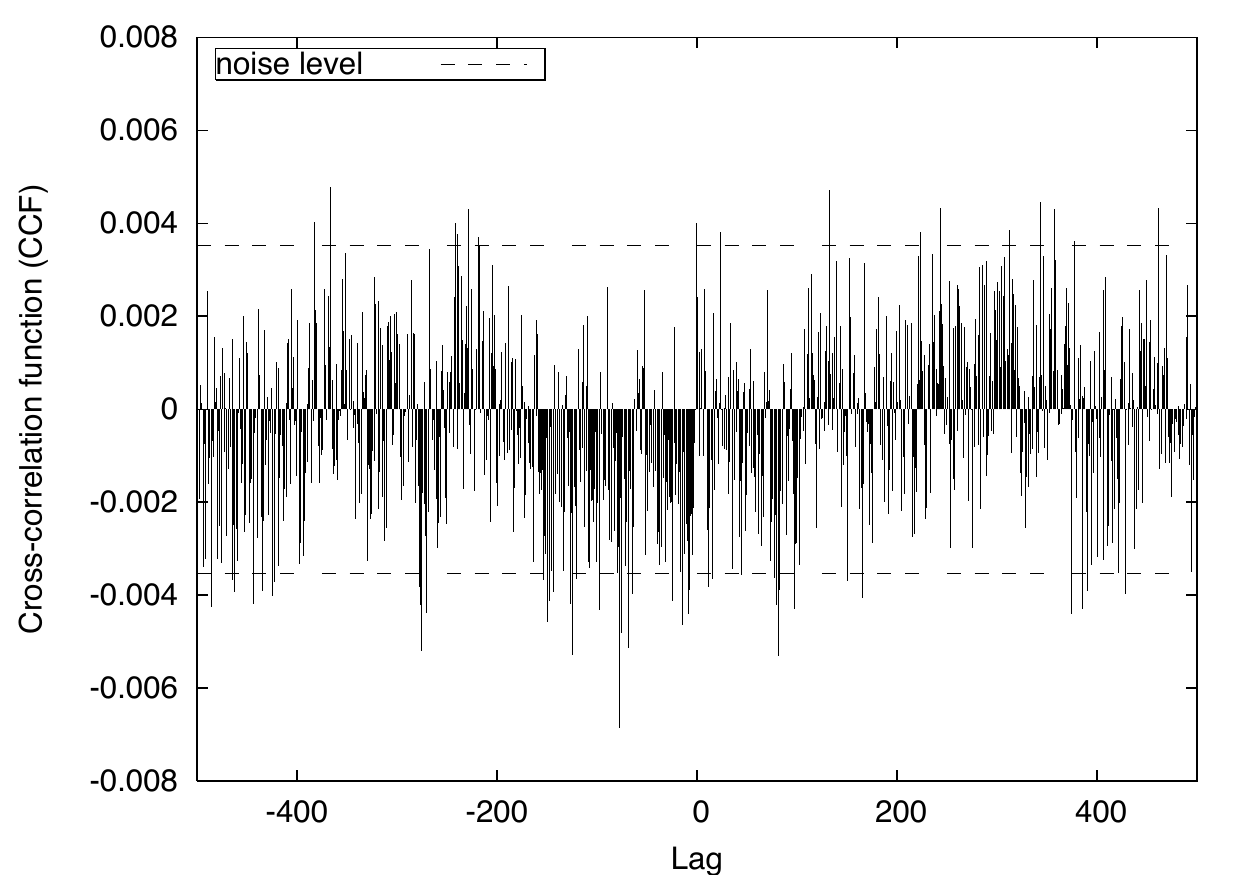}
}
\caption{ Lagged cross-correlation function of signs and absolute returns for the stock AZN.  The dashed line is the $2 \sigma$ standard error.}
\label{CCFsw_TTAZN}
\end{figure}
Finally we need to test the hypothesis of independence between signs and absolute returns.  This is not a simple task.  A first naive approach is to compute the cross-autocorrelation function between the two time series, as shown for AZN in Fig.~(\ref{CCFsw_TTAZN}).  For all the stocks in the sample the cross-correlations are practically negligible, always less than or comparable to  the noise level\footnote{This is computed as the $2 \sigma$-error, where $\sigma \equiv 1/\sqrt{N}$ and $N$ is the length of the series.} at any lag.  This might suggest that the assumption of independence between signs and sizes is a good approximation.  However, note the patterns in the autocorrelations in Fig.~(\ref{CCFsw_TTAZN}).  This suggests that even if the individual coefficients are small, there may be significant integrated effects, and in any case one must also worry about nonlinear interactions.  Later we will show that independence of the sign and size is not a good assumption.

\section{Estimating volatility}\label{VolatilityEstimation}

\subsection{Testing the model} \label{TestingTheModel}

Our goal is to test the validity of Eq.~(\ref{GRWVarFinal}).  To do this we divide the original time series into non-overlapping real time intervals of length $T=$ 15 minutes, 1 hour and 4 hours.  We measure the total price return $R_i$ during each interval $i$ and use $R_i^2$ for that interval as a proxy for its empirical squared volatility.  We compare this to the squared volatility prediction $\hat{V}_i$ based on Eq.~(\ref{GRWVarFinal}).  For each interval $i$ we estimate $\mu_s$, $\sigma_s$, $\mu_w$, $\sigma_w$, and count  the number of non-zero returns $n_i$.  In contrast, under our stationarity assumption $K_s(n_i)$ and $K_{w}(n_i)$ should not depend on which interval we choose.  This is fortunate because there is not enough data in an individual interval to get a statistically stable estimate.  Thus we estimate them for each stock using data from the entire time series.  Finally, we compute the ratio
\begin{equation}
\rho_i = \frac{R_i^2}{\hat{V}_i}.
\label{rho}
\end{equation}
If all the assumptions of the model were correct we should find that $\rho = 1$ to within statistical errors.

\begin{figure}[htbp!]
\resizebox{1.0\columnwidth}{!}{
	\includegraphics{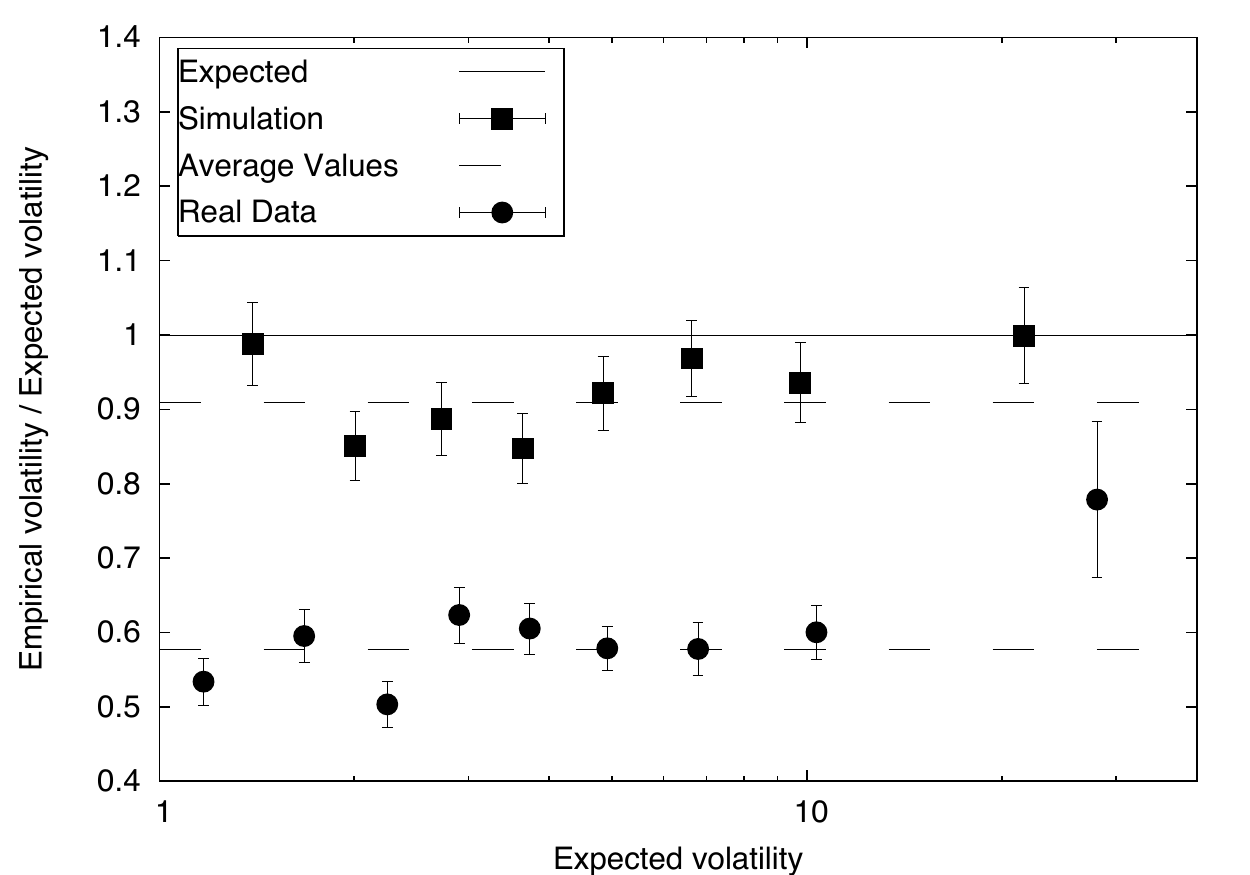}
}
\caption{The ratio $\rho$ of the empirical volatility and the expected volatility from Eq.~(\ref{rho}) as a function of the expected volatility.  The simulated data are shown as squares and the real data for AZN as circles.  Data are binned on the $x$-axis using quantiles with 10 bins and $538$ intervals per bin. The error bars represent standard errors.  The weighted mean value for the simulation is close to one, whereas the real data is closer to $0.6$.}
\label{EmpVolVSExpVolSimVSReal_AZN3600}
\end{figure}

As a reality check and to get a feeling for the expected statistical errors, we begin by testing this procedure on simulated data that is guaranteed to satisfy the assumptions of Section~\ref{assumptions}.  To make the test as realistic as possible we use the AZN original series of signs, and we generate a long-memory series of artificially generated absolute returns with a Hurst exponent $H = 0.7$ using a standard fractional Brownian motion generator.  This series explicitly differs from the real data in that it is log-normally distributed, and the sign and absolute return series are guaranteed to be independent of each other.  We ran the simulation on one hour intervals, i.e. we sample the artificial series in non-overlapping sub-intervals with same number of non-zero returns as in the one-hour sampling of the original AZN series.  Our results are reported in Fig.~(\ref{EmpVolVSExpVolSimVSReal_AZN3600}) where we show the average value of $\rho$ conditioned on the expected volatility.  
We consistently find $\rho \approx 0.9$.  Thus while our derived model gives a reasonable approximation, within ten percent of the correct answer, the predicted volatility is consistently slightly higher than the observed volatility.

We have performed extensive numerical experiments to understand the source of this bias that make it clear that the origin of this effect is statistical bias.  Because these data are highly skewed and display long-memory, when the variance is estimated with a finite number of data points the estimate of the volatility is systematically low.  This effect enters both for the empirical volatility itself (which we are measuring based on the price change across the whole interval, i.e. effectively with one point), and for the variance of the sizes of the returns.  Thus there is some cancellation, but the effect is more severe for the volatility, and hence the estimates tend to be low.  While we might be able to correct for this effect and improve the accuracy of our measurements, this is not trivial and the other effects that we observe later in this paper are sufficiently large that they dominate.


%

In contrast for real data the expected volatility is always significantly larger than the empirical volatility, as shown in Fig.~(\ref{EmpVolVSExpVolSimVSReal_AZN3600}) for AZN and reported for the other stocks in Table~(\ref{SummaryRatio}). For 1-hour sampling the ratio $\rho$ for AZN is close to $0.6$ and for the four stocks in our sample is in the range $\rho \in [0.59, 0.67]$.  As shown in Table~(\ref{SummaryRatio}) this overestimation also holds at all time scales, and gets worse as the time scale increases.  We conclude that our model consistently over-estimates the squared volatility by roughly $20\%$ or more at 15-minute time scales and more than $67\%$ at four hour time scales.

\begin{table*}[htdp]
\caption{Ratio of expected volatility to empirical volatility $\rho$ for real data and shuffling experiments.}
\begin{center}
\begin{tabular}{|l|l|c|c|c|c|}
\cline{3-6}  
\multicolumn{2} {l}{} & \multicolumn{1}{|c|}{AZN} & \multicolumn{1}{c|}{LLOY} & \multicolumn{1}{c|} {SHEL} & \multicolumn{1}{c|} {VOD} \\
\cline{3-6}\noalign{\smallskip}
\hline
& 15 min. &  $0.75\pm0.04$  & $0.84\pm0.05$ & $0.79\pm0.06$ & $0.71\pm0.03$\\
\cline{2-6}
Real data & 1 hour & $0.58\pm0.01$ & $0.66\pm 0.01$ & $0.63\pm0.01$ & $0.61\pm 0.01$\\
\cline{2-6}
 & 4 hours & $0.55\pm0.02$ & $0.57\pm0.03$ & $0.57\pm0.04$ & $0.59\pm0.03$\\
\hline
\hline
& Signs & $0.92\pm0.01$  & $0.94\pm0.02$ & $0.93\pm0.02$ & $0.93\pm0.02$\\
\cline{2-6}
Shuffling (1 h.) & Absolute returns &  $0.97\pm0.02$  &  $1.00\pm0.03$ & $0.97\pm0.02$ & $0.91\pm0.05$\\
\cline{2-6}
& Returns &  $1.02\pm0.02$  & $1.01\pm0.02$ & $1.02\pm0.02$ & $1.02\pm0.03$\\
\hline
\hline
Block & Signs and Sizes &  $0.92\pm0.02$  & $0.95\pm0.02$ &  $0.96\pm0.02$   & $ 0.93\pm0.02$\\
\cline{2-6}
shuffling (1 h.) & Returns &  $0.68\pm0.02$  &  $0.75\pm0.02$ & $0.71\pm0.01$  & $0.73\pm0.03$\\
\hline
\end{tabular}
\end{center}
\label{SummaryRatio}
\end{table*}

\subsection{Shuffling experiments}

To understand why the model of Eq.~(\ref{GRWVarFinal}) fails we perform a series of shuffling experiments at the one hour time scale.  In each case we randomly rearrange the order of a given component of the real data while preserving everything else.
\begin{enumerate}
\item
{\it Signs}. We randomly shuffle the sign time series.  This destroys the autocorrelation of the signs and any cross-correlation between signs and absolute returns, but preserves the autocorrelation of the absolute returns.
\item
{\it Absolute returns}.  We shuffle the absolute return time series.  This destroys the autocorrelation of absolute returns and any cross-correlation between signs and absolute returns, but preserves the autocorrelation of the signs.
\item
{\it Returns}.  We shuffle returns, i.e. we shuffle both signs and absolute returns together, using the same permutation.   This destroys the autocorrelation of both signs and sizes, and at the same time preserves their contemporaneous cross-correlation while destroying any lagged cross-autocorrelations.
\end{enumerate}
In each case we measure $\rho$ just as we did for the real data.   

The results of these experiments are shown in Table~(\ref{SummaryRatio}).  We see that when we shuffle signs we observe $\rho \approx 0.93$ fairly consistently for each stock.  This is smaller than one, but much larger than the $\rho$ observed for real data.  This result is consistent with the bias we observed earlier in our benchmark  simulating long-range correlated absolute returns.
However, this effect is much too small to explain the large discrepancy with the real data -- there must be another, much larger effect in the real series of absolute returns.

In contrast, when we shuffle the absolute returns or the returns, we observe $\rho \approx 1$ in almost every case\footnote{The exception is when we shuffle absolute returns for Vodafone we observe $\rho = 0.91 \pm 0.05$.  We don't know whether this implies that our error bars (based on standard errors) are too optimistic or whether there is some effect that makes Vodafone different from the other stocks.}  From this we conclude that the underestimation we observe is neither caused by the autocorrelation of signs nor is it caused by the contemporaneous cross correlation of signs and absolute returns.

\begin{figure}[htbp]
\resizebox{1.0\columnwidth}{!}{
 	\includegraphics{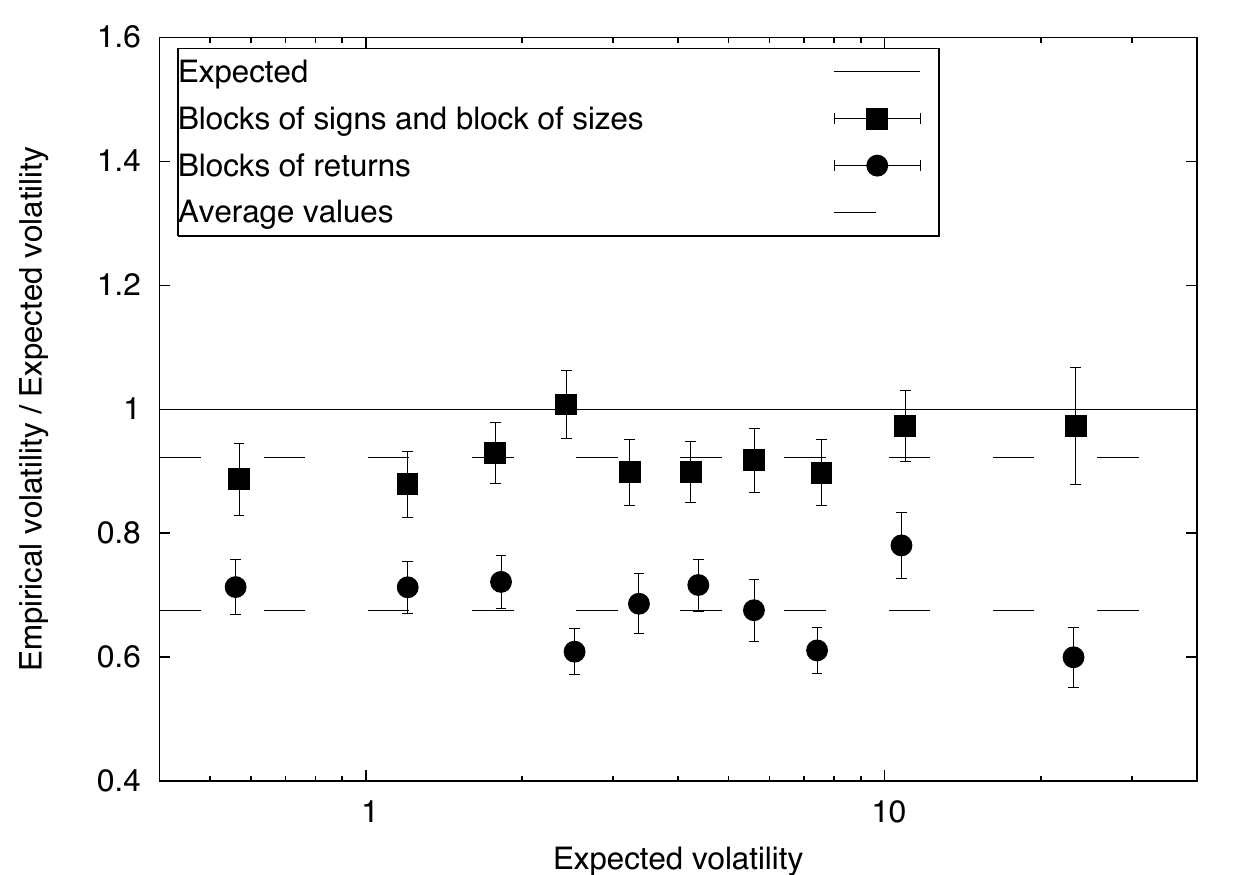}
}
\caption{{\bf Block shuffling experiments for AZN}. We compare shuffling blocks of returns to shuffling blocks of absolute returns and signs separately using blocks of length $L=60$.  Data are binned along the $x$-axis based on their expected volatility in 10 bins with 538 one hour intervals per bin.  The ratio $\rho$ plotted on the vertical axis indicates whether Eq.~(\ref{GRWVarFinal}) correctly predicts the volatility for the shuffled data sets in each range of the expected volatility; error bars are standard errors.  A horizontal black line at $y=1$ is shown for comparison.  Circles are for shuffling blocks of returns, triangles for shuffling signs and absolute values separately, and the dashed lines are the mean values of each.  Shuffling signs and absolute returns separately destroys their lagged cross-correlation, and results in correct estimates, while shuffling returns produces a similar over-estimation to that observed for the real data.  This supports our hypothesis that a subtle correlation between absolute returns and signs causes the overestimation for real data.}
\label{EmpVol_vs_ExpVol_BLOCKSHUFF}
\end{figure}

These experiments suggest that the main cause of the over-estimation of volatility is the lagged relationship between signs and absolute returns\footnote{This result is in line with that obtained by Weber \cite{Weber07}, who noticed that the size of the largest price changes is over-estimated if one assumes that signs and absolute returns are independent. Moreover, this result is in some way similar to the leverage effect\cite{Bouchaud01b}, even if it is usually observed on daily or even weekly time scales, while our result holds at individual transaction time scale.}.  To test this more explicitly we perform block shuffling experiments, in which we randomly interchange the order of blocks of length $L$ while leaving everything the same within each block.  We performed two different tests:  
\begin{enumerate}
\item
{\it Blocks of signs and sizes separately}. We shuffle blocks of signs and absolute returns separately.   This preserves the individual autocorrelation structures up to the block size, but destroys any cross correlation between the signs and the absolute returns\footnote{When we shuffle blocks of signs and absolute returns separately we use different block boundaries for each.  Thus a sign for a given return is typically matched with a different absolute return.}.
\item
{\it Blocks of returns}.  We shuffle blocks of returns, keeping the same ordering of signs and sizes within the block.  This preserves all the autocorrelations and all the lagged cross correlations between signs and absolute returns up to the size of the blocks.
\end{enumerate}
The results for blocks of length $60$ are summarized for AZN in Fig.~(\ref{EmpVol_vs_ExpVol_BLOCKSHUFF}) and for all stocks in Table~\ref{SummaryRatio}.  
In the experiments where we shuffle blocks of signs and sizes we find $\rho \in [0.92, 0.96]$. As we previously observed when shuffling signs alone with block lengths of one, $\rho$ is slightly smaller than one, consistent with simulations reported in Sec.~\ref{TestingTheModel} Thus, even if there is a small tendency for Eq.~(\ref{GRWVarFinal}) to overestimate volatility and for the proxy $R_n^2$ to underestimate volatility, our simulations show that this effect is small and is not  sufficient to explain the discrepancy observed in the real data.  This once again suggests that preserving the relationship between signs and sizes is important.

\begin{figure}[htbp]
\resizebox{1.0\columnwidth}{!}{
	\includegraphics{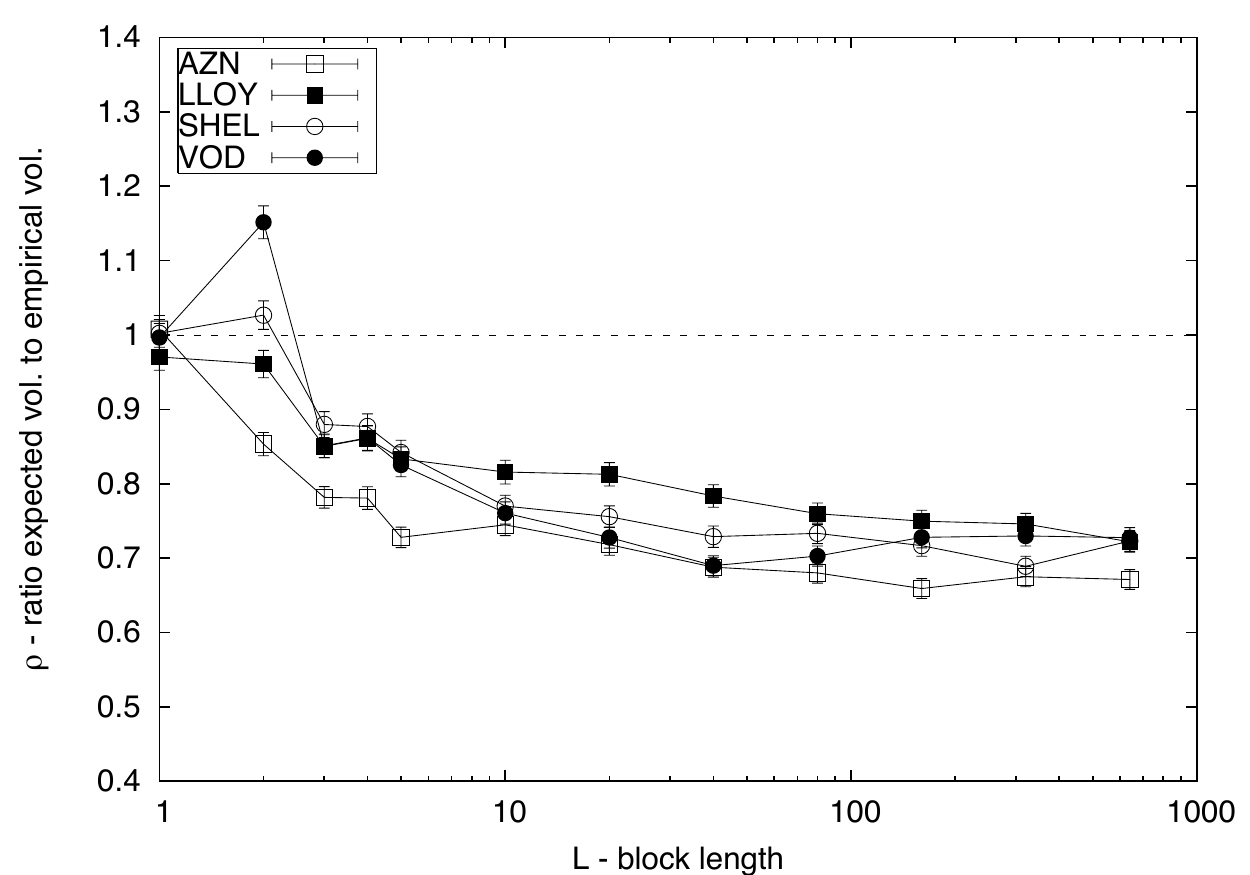}
}
\caption{{\bf Dependence of $\rho$ on block size for return shuffling experiments for one hour intervals}.  In the original time series blocks of length $L$ are shuffled, preserving the ordering of signs and absolute returns within each block.  The ratio $\rho$, which measures the amount by which Eq.~(\ref{GRWVarFinal}) over-estimates volatility, is plotted as a function of the block size $L$ for each of the four stocks in our sample.   This shows that as the block size increases the estimates decrease fairly steadily toward the observed values for the real data.}
\label{rhoVsBlockLength}
\end{figure}

In contrast, when we test this directly by shuffling blocks of returns while preserving the relationship between signs and sizes, we find a significant overestimation of the volatility with $\rho \sim 0.7$.  This value is significantly smaller than one, making it clear that we have captured most of the effect, but it is still larger than the value $\rho \sim 0.6$ that we observed for the real data.  We believe that this is because the block length $L = 60$ is not long enough.  To test whether this is the case in Fig.~(\ref{rhoVsBlockLength}) we plot the estimated value of $\rho$ for a block return shuffling experiment for each stock as a function of block length.
As for our previous experiments with $L = 1$, we observe $\rho \approx 1$.  As $L$ increases $\rho$ decreases fairly steadily in every case.  However, this decrease is fairly slow, and at the maximum block size length $L = 600$ it has still not decreased to the low value $\rho \sim 0.6$ observed for the full sample.  We believe this is because the maximum block length, which is limited by our ability to obtain good statistical sampling, is still not long enough.  This suggests the time scale of the  cross-correlations between signs and absolute returns is very long.

 
\section{Discussion and conclusions}\label{Conclusions} 

Under the assumptions given in Section~\ref{assumptions} we have derived a formula for volatility under a simple generalized random walk model.  For a time interval of any given length, this formula relates volatility to simple properties of the underlying random walk, in particular the number of non-zero returns, and the mean and variance of the signs and absolute values of returns, as well as their integrated autocorrelation.

We find that this formula consistently overestimates volatility.  We have shown that the main reason for this is because our formula assumes that return signs and return sizes are independent.  In contrast, for the real data there are long-range correlations, which are small at any given time lag but large when integrated over long time scales.   This effect is quite large:  The overestimate is roughly $67\%$ for one hour intervals, and even more for four hour intervals.  

These results are surprising because they indicate that the volatility is reduced by almost a half due to a subtle long range interaction between the signs of returns and their sizes.  This is particularly surprising because it involves signs of returns and not the signs of transactions.  The signs of transactions form a long-memory process while the signs of returns do not.  Thus the evidence seems to indicate that there is a very-long range interaction between return signs and sizes, even though return signs themselves do not show long-memory properties.

We believe that this interaction is closely related to the interaction that takes place between the transaction signs and returns as studied in references \cite{Bouchaud04,Lillo03c,Bouchaud04b,Farmer06}, but at this point we have not been able to show this.  Intuitively this can be seen as follows:  Because of the long-memory properties of transactions, which make their signs highly predictable, returns must compensate so that they are not equally predictable.  One way to make this happen, as stressed by Bouchaud et al., is that price impacts are temporary, i.e. when transactions happen prices change but this change decays slowly with time.  Alternatively, as stressed by Lillo and Farmer, price changes can have a permanent component, but this component varies based on the predictability of transaction signs:  When a future transaction is very likely to be a buy, the size of buy returns is much larger than the size of sell returns.  These two approaches have been shown to be equivalent \cite{Gerig07}.  In either case it suggests a reduction of volatility relative to what one would expect under an unconditional permanent impact model such as the one we have developed here.  In a future paper we hope to show that this is indeed the reason for the missing component of volatility, or alternatively provide a better explanation.

\paragraph*{}

\section*{Acknowledgments}
{\small We would like to thank Miguel Virasoro and Andrea De Martino for useful comments. FL acknowledges support from the research project MIUR 449/97 ``High frequency dynamics in financial markets" and from the European Union STREP project n. 012911 ``Human behavior through dynamics of complex social networks: an interdisciplinary approach.''.   GL and JDF acknowledge support from Barclays Bank, Bill Miller and NSF grant HSD-0624351.  Any opinions, findings and conclusions or recommendations expressed in this material are those of the authors and do not necessarily reflect the views of the National Science Foundation.}

\bibliographystyle{plain}
\bibliography{jdf}

\end{document}